\documentclass[aps,prl,reprint,showpacs,superscriptaddress,
nofootinbib,nofootinbib
]{revtex4-1}
\usepackage{graphicx}
\usepackage{amsfonts}
\usepackage{amsmath}
\usepackage{amssymb}
\usepackage{txfonts} 
\usepackage{times}
\usepackage{subfigure}
\usepackage{textcomp}
\usepackage{color}
\definecolor{gray}{gray}{0.8}
\usepackage{dcolumn}
\usepackage{bm}
\usepackage{hyperref}

\newcommand{\beq}{\begin{equation}}     \newcommand{\eeq}{\end{equation}}
\newcommand{\beqa}{\begin{eqnarray}}    \newcommand{\eeqa}{\end{eqnarray}}
\newcommand{\bde}{\begin{description}}  \newcommand{\ede}{\end{description}}
\newcommand{\ben}{\begin{enumerate}}    \newcommand{\een}{\end{enumerate}}
\newcommand{\ind}{\indent }

\newcommand{\la}{\langle}               \newcommand{\ra}{\rangle}

\newcommand{\kT}{{k_{\rm B}T} }

\newcommand{\intII}{{\int^{+\infty}_{-\infty}}}
\newcommand{\intOI}{{\int^{+\infty}_{0}}}

\newcommand{\eqn}[1]{\beq{ #1 }\eeq}

\newcommand{\inv}[1]{{\frac{1}{#1}}}

\newcommand{\inRbracket}[1]{{\left({#1}\right)}}

\newcommand{\inCbracket}[1]{{\left\{{#1}\right\}}}

\newcommand{\inAverage}[1]{{\left\la{#1}\right\ra}}

\newcommand{\blue}[1]{{\color{blue}{#1}\normalcolor}}

\begin{document}
\title{Non-equilibrium statistical mechanics of the heat bath for two Brownian particles}
 \author{Caterina De Bacco} \affiliation{Laboratoire de Physique Th\`{e}orique et Mod\`{e}les Statistiques, CNRS et Universit\'e  Paris-Sud 11, UMR8626, B\^at.100, 91405 Orsay Cedex, France} 
 \author{Fulvio Baldovin} 
 \affiliation{Dipartimento di Fisica e Astronomia G. Galilei and Sezione INFN, Universit\`{a} di Padova, Via Marzolo 8, I-35100 Padova, Italy}
\author{Enzo Orlandini} 
 \affiliation{Dipartimento di Fisica e Astronomia G. Galilei and Sezione INFN, Universit\`{a} di Padova, Via Marzolo 8, I-35100 Padova, Italy}
 \author{Ken Sekimoto} \affiliation{Mati\`{e}res et Syst\`{e}mes Complexes, CNRS-UMR7057, Universit\'e   Paris-Diderot, 75205 Paris, France}  \affiliation{Gulliver, CNRS-UMR7083, ESPCI, 75231 Paris, France}
\begin{abstract}

{We propose a new look at the heat bath for two Brownian particles, 
in which the heat bath as a `system' is both {perturbed and sensed} by
the Brownian particles. Non-local thermal fluctuation give rise to
bath-mediated static forces between the particles.  Based on the
general sum-rule of the linear response theory, we derive an explicit
relation linking these forces to the friction kernel describing the
particles' dynamics.  The relation is analytically confirmed in the
case of two solvable models and could be experimentally challenged.  Our
results point out that the inclusion of the environment as a part of
the whole system is important for micron- or nano-scale physics. } 
\end{abstract}

\maketitle

\ind\blue{\em {Introduction ---}} 
{Known as the thermal Casimir interactions \cite{Tcasimir1978}  or the Asakura-Oosawa interactions \cite{asakura1954}, a fluctuating environment can mediate static forces between the objects constituting its borders.  
Through a unique combination of the generalized Langevin equation and the linear response theory, we uncover a link between such interactions and the correlated Brownian motions with memory, 
both of which reflect the spatiotemporal non-locality of the heat bath.}

\begin{figure}[h]{}
\centering{\includegraphics[width=2.0in]{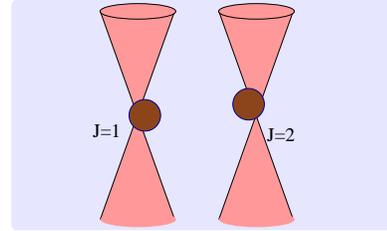}}
\caption{\label{fig:general}
Two Brownian particles (filled disks, $J=1$ and $J=2$)
are trapped by an external potential, such as through optical traps (vertical cones), and interact through both the direct and the heat bath-mediated interactions.}
\end{figure}

The more fine details of Brownian motion are experimentally revealed, the more deviations from the idealized Wiener process are found  (see, for example, \cite{Hinch-exp3}).
When two Brownian particles are trapped close to each other in a heat bath 
(see \hyperref[fig:general]{Fig.\ref{fig:general}}), the random forces on those objects are no more 
independent noises but should be correlated.
Based on the projection methods \cite{Mori-formula,Zwanzig60,Zwanzig61} we expect the generalized Langevin equations to apply
\cite{Kawasaki73,zwanzig73,ZwanzigBook,Ciccotti-Ryckaert-JSP1980}:
\beq
\label{eq:Pdot}
M_J	 \frac{d^2 X\!_J(t)}{dt^2}
\!=-\frac{\partial U}{\partial X_J}
- \!\!  \sum_{J'=1}^2  \int^t_0  
 \!\!\! K_{J,J'}(t-\tau)\frac{dX_{J'}(\tau)}{d\tau}d\tau +\epsilon_J(t),
\eeq 
where $X_J$ ($J=1\mbox{ and }2$) are the position of the Brownian particles with the mass being $M_J,$ and $K_{J,J'}(s)$ and $\epsilon_J(t)$ are, respectively, the friction kernel  and the random force.
 $U(X_1,X_2)$ is the static interaction potential between the Brownian particles.
If the environment of the Brownian particles at the initial time $t=0$ is in canonical equilibrium at temperature $T$,  the noise and the frictional kernel should satisfy the fluctuation-dissipation (FD) relation of the second kind with the Onsager symmetries \cite{Kawasaki73,Zwanzig75}: 
\eqn{\label{eq:FD2}
\la \epsilon_J(t)\epsilon_{J'}(t')\ra =\kT\,K_{J,J'}(t-t'),
}
\eqn{\label{eq:Onsager} 
K_{J,J'}(s)=K_{J',J}(s)=K_{J,J'}(-s),
}
where $J$ and $J'$ are either 1 or 2 independently.
%
{This model  (\ref{eq:Pdot}) is a pivotal benchmark model for the correlated Brownian motion, 
although the actual Brownian motions could be more complicated (see, for example, \cite{nonMarkov-Bocquet-JSP1997,Hinch-exp3}).} 
But ``the physical meaning of the random force autocorrelation function is in this case far from clear...'' even now and  ``A proper derivation of the effective potential could be of great help in clarifying this last point''  \cite{Ciccotti-Ryckaert-JSP1980}.
In addition to the bare potential $U_0(X_1,X_2)$ independent of the heat bath, 
 the potential $U,$ {which is in fact the free energy as function of $X_J$},  may contain a bath-mediated interaction potential {$U_{{b}}(X_1,X_2)$}  so that
 \eqn{U(X_1,X_2)=U_0(X_1,X_2)+{U_{{b}}(X_1,X_2)}.}

In this Letter we propose the relation
\eqn{\label{eq:main} 
K_{1,2}(0)=-\frac{\partial }{\partial X_1}\frac{\partial }{\partial X_2}U_{{b}}(X_1,X_2), 
}
where {the both sides of this relation should be evaluated at the equilibrium positions of the Brownian particles, $X_J=\la X_J\ra_{\rm eq}.$}
This relation implies that the bath-mediated static interaction 
is always correlated with  the frictional one. Our approach is to regard the heat bath as the weakly non-equilibrium system which is both perturbed and sensed by the 
mesoscopic Brownian particles.
From this point of view (\ref{eq:main}) is deduced from so called `general sum-rule theorem' \cite{JPSJ.12.570} 
of the linear response theory of non-equilibrium statistical mechanics \cite{Kubo}. 
While the FD relation of the second kind (\ref{eq:FD2}) is well known as an outcome of this theory, the other aspects have not been fully explored.
Below we give a general argument supporting (\ref{eq:main}), and then give two analytically solvable examples for which the claim holds exactly. 
 
\ind\blue{\em {General argument ---}} 
While the spatial dimensionality is not restrictive in the following argument, 
 we will use the notations as if the space were one-dimensional. 
Suppose we observe the force $F_{1,2}$ on the $J=1$ particle
 as we move the $J=2$ particle from $\la X_2\ra_{\rm eq}$ at $t=-\infty$ to $X_2(t)$ at $t$.
Due to the small perturbation $X_2(t)-\la X_2\ra_{\rm eq}$, the average force at that time, $\la F_{1,2}\ra_t,$ is deviated from its its equilibrium value, $\la F_{1,2}\ra_{\rm eq}.$ 
The linear response theory relates these two through the response function, $\Phi_{1,2}(s)$ as 
\eqn{\label{eq:FPhi}
\inAverage{F_{1,2}}_{t}\!-\inAverage{F_{1,2}}_{\rm eq.}\!\!\!
=
\!\!\int^t_{-\infty}\!\!\!\Phi_{1,2}(t-\tau)\,(X_2(\tau)-\la X_2\ra_{\rm eq})d\tau.
}
(Within the linear theory the force is always measured at $X_1=\la X_1\ra_{\rm eq}$.) 
The complex admittance $\chi_{1,2}(\omega)
=\chi\,'_{1,2}(\omega)+i\chi^{\,\prime\prime}_{1,2}(\omega)$ is defined as 
the Fourier-Laplace transformation of $\Phi_{1,2}(s)$ : 
\eqn{\label{eq:chidef}  
\chi_{1,2}(\omega)=\intOI e^{i\omega s-\varepsilon s}\Phi_{1,2}(s)ds,
}
where $\varepsilon$ is a positive infinitesimal number (i.e., $+0$). 
If $\chi_{1,2}(\infty)=0$, which is the case in the present , 
the causality of 
$\Phi_{1,2}(t),$ or the analyticity of $\chi_{1,2}(\omega)$ in the upper half complex plane of $\omega$, impose the general sum rule \cite{JPSJ.12.570},
\eqn{\label{eq:sumrule3} 
{\cal P}\intII \frac{\chi_{1,2}^{\,\prime\prime}(\omega)}{\omega}\frac{d\omega}{\pi}=\chi\,'_{1,2}(0),
}
where $\cal P$ on the left hand side (l.h.s.) denotes to take the principal value of the integral across $\omega=0$. 
The significance of (\ref{eq:sumrule3}) is that it relates the dissipative quantity (l.h.s.) and the reversible static response (right hand side (r.h.s.)) of the system.

Now we suppose, along the thought of Onsager's mean regression hypothesis \cite{Onsager2}, that the response of the heat bath to the fluctuating Brownian particles, which underlies (\ref{eq:Pdot}), 
 is essentially the same as the response to externally
specified perturbations described by (\ref{eq:FPhi}). Thus the comparison of (\ref{eq:FPhi}) with (\ref{eq:Pdot}) gives 
\eqn{\label{eq:PhidKdt}
\Phi_{1,2}(t)=-\frac{dK_{1,2}(t)}{dt},}
or, in other words, $K_{1,2}$ is the relaxation function corresponding to $\Phi_{1,2}.$
With this linkage between the Langevin description and the linear response theory, 
the static reversible response $\chi'_{1,2}(0)$ of 
the force $\la F_{1,2}\ra-\la F_{1,2}\ra_{\rm eq}$ to the static displacement
 $X_2-\la X_2\ra_{\rm eq}$ can be identified with the r.h.s. of (\ref{eq:main}).
As for the l.h.s. of (\ref{eq:sumrule3}), we can show by  (\ref{eq:PhidKdt}) and (\ref{eq:chidef}) that it is equal to $K_{1,2}(0).$ 
The argument presented here is to be tested both analytically/numerically and experimentally. At least for the two models presented below the claim (\ref{eq:main}) is analytically confirmed.

\begin{figure}[h]
\centering{\includegraphics[width=3.in]{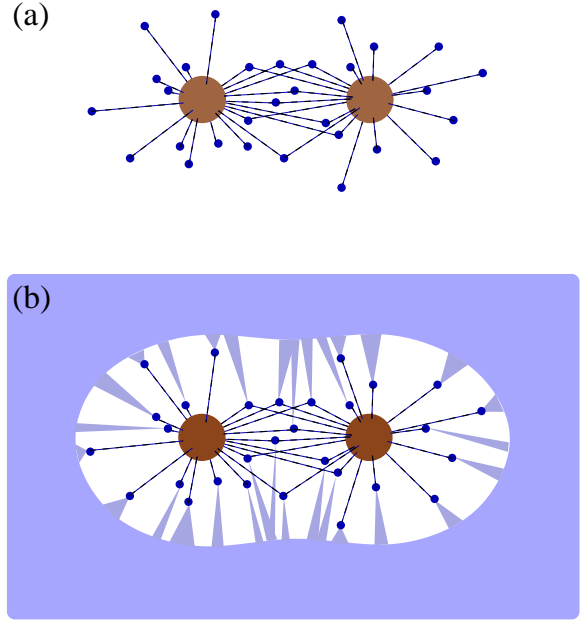}}
\caption{
{\it (a)} Hamiltonian model of two Brownian particles which is analytically solvable for one dimensional space with harmonic coupling. Each light mass particle (thick dot) is linked to at least one of the Brownian particles (filled disks) with Hookean springs (dashed lines). {\it (b)} Langevin model of two Brownian particles. Unlike the Hamiltonian model, each light mass particle receives the random force and frictional force from the background (shaded zone) and its inertia is ignored.
}
\label{fig:2} 
\end{figure}

\ind\blue{\em {Solvable model I: Hamiltonian system---}} 
As the first example that confirms the relation (\ref{eq:main}) we take up a Hamiltonian model 
inspired by the classic model of Zwanzig \cite{zwanzig73}, see Fig.~\ref{fig:2}(a). 
Instead of a single Brownian particle \cite{zwanzig73} we put the two Brownian particles
with masses $M_J$ ($J=1,2$) which interact with the `bath' consisting of light mass  `gas' particles.
While Fig.~\ref{fig:2}(a) gives the general idea, the solvable model is limited to the one-dimensional space. 
Each gas particle, e.g. $i$-{th} one, has a mass $m_i$ ($\ll M_J$) and is linked to at least one of the Brownian particles, $J=1\mbox{ or }2$, through Hookean springs of the spring constant $m_i\omega_{i,J}^2 (>0)$ and the natural length, $\ell_{i,J}$. 
In Fig.~\ref{fig:2}(a) these links are represented by the dashed lines. 
The Hamiltonian of this purely mechanical model consists of three parts, 
$H=H_B+H_b+H_{bB}$, with
\eqn{\label{eq:H0}
H_B=\frac{P_1^2}{2M_1}+\frac{P_2^2}{2M_2}+U_0(X_1,X_2),
}
\eqn{\label{eq:Hb}
H_b=\sum_i \frac{p_i^2}{2m_i}, \quad
H_{bB}=\sum_i \frac{m_i}{2}\sum_{J=1}^2 \omega_{i,J}^2 (q_i-X_J-\ell_{i,J})^2,}
where the pairs $(X_J,P_J=$$M_J\,dX_J/dt)$ and $(x_i,p_i=$$m_i\,dx_i/dt)$ denote, respectively, the positions and momenta of the heavy ($J$) and light ($i$) particles. 
The Brownian particles obey the following dynamics :
\eqn{\label{eq:withT}
M_J\frac{d^2 X_J}{dt^2}
= -\frac{\partial U_0}{\partial X_J}
+\sum_i m_i  \omega_{i,J}^2(q_i-X_J-\ell_{i,J}). 
}

Given the initial values of $(q_i,p_i)$ at $t=0$, the Hamilton equations for $(q_i(t),p_i(t)),$ which reads
\eqn{\label{eq:qdots} 
m_i\frac{d^2 q_i}{dt^2} = -m_i\sum_{J=1}^2\omega_{i,J}^2 (q_i-X_J(t)-\ell_{i,J}),
}   
can be solved in supposing that the histories of $X_J(s)$ ($J=1\mbox{ and }2$) for $0\le s\le t$ are given. 
In order to assure the compatibility with the initial canonical equilibrium of the heat bath, 
we assume the vanishing initial velocity for the Brownian particles, $\left.dX_J/dt\right|_{t=0}=0.$ 
Substituting each $q_i$ in (\ref{eq:withT}) by its formal solution thus obtained, 
the dynamics of $X_J(t)$ is rigorously reduced to  (\ref{eq:Pdot}), where 
 the friction kernels $K_{J,J'}(s)$ are 
\eqn{\label{eq:KJJ}
K_{J,J'}(s)=\sum_i \frac{m_i \omega_{i,J}^2\omega_{i,J'}^2}{\tilde{\omega}_i^2}\cos(\tilde{\omega}_i s),
}
and the noise term $\epsilon_J(t)$ is  
\eqn{\label{eq:Noise}
\epsilon_J(t)\equiv \sum_i m_i\omega_{i,J}^2\inCbracket{
\tilde{q}_i(0) \cos(\tilde{\omega}_i t)
+\frac{d\tilde{q}_i(0)}{dt}\,
\frac{\sin(\tilde{\omega}_i t)}{\tilde{\omega}_i}
},}
with  $\tilde{\omega}_i^2\equiv  {\omega_{i,1}^2+\omega_{i,2}^2}$ and  
\eqn{\label{eq:qtilde}
\tilde{q}_i(t)
\equiv q_i(t)-\sum_{J=1}^2\frac{\omega_{i,J}^2}{\tilde{\omega}_i}[\ell_{i,J}+X_J(t)].
}
{To our knowledge this is the first concrete model that demonstrates (\ref{eq:Pdot}).}
Only those gas particles linked to the both Brownian particles satisfy $\omega_{i,1}^2\omega_{i,2}^2>0$ and contribute to $K_{1,2}(s)$. 
While the generalized Langevin form (\ref{eq:Pdot}) holds for an individual realization without any ensemble average, the statistics of $\epsilon_J(t)$ must be specified. We assume that at $t=0$ the bath variables $\tilde{q}_i(0)$ and $\tilde{p}_i(0)\, (=$${p}_i(0)$  because we defined $\left.dX_J/dt\right|_{t=0}=0$) 
belong to the canonical ensemble of a temperature $T$ with the weight $\propto \exp(-(H_b+H_{bB})/\kT).$
Then the noises $\epsilon_J(t)$ satisfy the FD relation of the second kind (\ref{eq:FD2}). and the Onsager symmetries (\ref{eq:Onsager}).

In this solvable model, the heat bath-mediated static potential $U_{{b}}$ which supplements $U_0$ to make $U
=U_0+U_{{b}}$ is found to be
\eqn{\label{eq:Ub}
U_{{b}}(X_1-X_2)=\frac{k_{{b}}}{2}(X_1-X_2{-}{L_{{b}}})^2,}
where
\eqn{\label{eq:kbLb}
k_{{b}}=
\sum_i \frac{m_i \omega_{i,1}^2\omega_{i,2}^2}{\tilde{\omega}_i^2} , \quad
L_{{b}}=
\inv{{k_{{b}}}}\sum_i\frac{m_i \omega_{i,1}^2\omega_{i,2}^2
(\ell_{i,1}-\ell_{i,2})}{\tilde{\omega}_i^2}.
}
Note that $U_{{b}}$ depends on $X_1$ and $X_2$ only through $X_1-X_2$, that is, it possesses the translational symmetry (see later).
While this form appears in the course of deriving (\ref{eq:Pdot}), its origin can be simply understood from the following identity:
\eqn{
\label{eq:Hb-bis}
H_{bB}  
=\sum_i \frac{m_i\tilde{\omega}_i^2}{2}\tilde{q}_i^2
+  U_{{b}}(X_1-X_2).
}
Finally, our claim (\ref{eq:main}) is confirmed by (\ref{eq:KJJ}) for $K_{1,2}(0)$ and by (\ref{eq:Ub}) and 
(\ref{eq:kbLb}) for the $U_{{b}}^{\,\prime\prime}(X)=k_{{b}}.$ 
In the standard language of the linear response theory, the `displacement' $A$ conjugate to the external parameter  $X_2(t)-\la X_2\ra_{\rm eq}$ is $A=\sum_i m_i \omega_{i,2}^2(q_i-X_2-\ell_{i,2})$ and the flux as the response is $B=\sum_i m_i \omega_{i,1}^2(q_i-X_1-\ell_{i,1})$ 
\cite{Kubo}. Direct calculation gives $\chi_{1,2}(\omega)=\sum_i(m_i\omega_{i,1}^2\omega_{i,2}^2)/[\tilde{\omega}_i^2-(\omega+i\varepsilon)^2].$

A remark is in order about the {translational symmetry of $U_{{b}}(X)$}.
In the original Zwanzig model \cite{zwanzig73},
the factor corresponding to $q_i-X_J-\ell_{i,J}$ in (\ref{eq:Hb}) was $q_i-c_iX_J$ with an arbitrary constant $c_i$ {and the} natural length $\ell_{i,J}$  set to be zero arbitrarily.
In order that the momentum in the heat bath is locally conserved around two Brownian particles,
we needed to set $c_i=1$ and explicitly introduce the natural length $\ell_{i,J},$ especially for those gas particles
which are coupled to the both Brownian particles, i.e. with $\omega_{i,1}^2\omega_{i,2}^2> 0.$ 
We note that the so-called dissipative particle dynamics modeling  \cite{DPD-originEPL1992,DPD-secondEPL1995,LoweEPL1999} also respects the local momentum conservation.

\ind\blue{\em {Solvable model II: Langevin system.---}}
The second example that confirms the relation (\ref{eq:main}) is constructed by modifying the first one, see Fig.~\ref{fig:2}(b).
There, we replace the Hamiltonian evolution of each light mass particle (\ref{eq:qdots}) by 
the over-damped stochastic evolution governed by the Langevin equation; 
\eqn{\label{eq:qLangevin}
0 = -\gamma_i \frac{dq_i}{dt}+\xi_i(t) 
-m_i\sum_{J=1}^2\omega_{i,J}^2 (q_i-\ell_{i,J}-X_J(t)),
}
where $\gamma_i$ is the friction constant with which 
the $i$-th gas particle is coupled to a `outer'-heat bath of the temperature $T.$\,
$\xi_i(t)$ is the Gaussian white random force from the outer-heat bath  
 obeying $\la \xi_i(t)\ra=0,$ and $\la \xi_i(t)\xi_{i'}(t')\ra=$$2\gamma_i$$\kT$$\delta(t-t')\delta_{i,i'}.$
This outer-heat bath may represent those degrees of freedom of the whole heat bath which are not directly coupled to the Brownian particles, while the variables $(q_i,p_i)$ represent those freedom of our primary interest as the `system'.
(Similar idea has already been proposed in different contexts, see \cite{LNP} \S 6.3 and \S 7.1, and also \cite{carnotsteng,puglisi-cage-EPL2010,Komatsuzaki2011a}.)
Integrating (\ref{eq:qLangevin}) for $q_i(t)$ and substituting the result into the r.h.s. of (\ref{eq:withT}), we again obtain (\ref{eq:Pdot}) and (\ref{eq:FD2}) with the same
bath-mediated static potential as before, i.e., $U_{{b}}$ defined by (\ref{eq:Ub}) and (\ref{eq:kbLb}).
({In this over-damped model, $m_i \omega_{i,J}^2$ simply represents the spring constant between the $i$-th light mass and the $J$-th Brownian particle.})
The friction kernel and the noise term of the present model are, however, different: instead of (\ref{eq:KJJ}) and (\ref{eq:Noise}), they read, respectively,
\beq \label{dq:KJJiso}
K_{J,J'}(s)=\sum_i \frac{m_i \omega_{i,J}^2\omega_{i,J'}^2}{\tilde{\omega}_i^2}
e^{-\frac{|s|}{\tau_i}},
\eeq
\beq
\epsilon_J(t)=\sum_i m_i\omega_{i,J}^2 \int_0^\infty \frac{e^{-\frac{s}{\tau_i}}}{\gamma_i}\xi_i(t-s)ds,
\eeq
where $\tau_i=\gamma_i/(m_i\tilde{\omega}_i^2).$
Because the form of $K_{1,2}(0)$ as well as $U_{{b}}(X)$ are unchanged from the first model,
our claim (\ref{eq:main}) is again confirmed.

\ind\blue{\em {Discussion : Implication of (\ref{eq:main}) ---} } 
Being consistent with this relation, no bath-mediated interactions appeared
 in the phenomenological approaches \cite{Ermak1978,Ferderhof1991,Dufresne2000} where the Stokesian fluid model is supplemented by the thermal random forces satisfying the FD relation, 
because the bath had no memory.

The above solvable models, though being artificial, represent certain non-local aspects of the more realistic heat baths.
The cross frictional kernel $K_{1,2}(s)$ and the bath mediated potential $U_{{b}}(X)$ are  generated by those microscopic degrees of freedom which couple to both the Brownian particles. 
This picture is reminiscent of the quantum system interacting with electromagnetic fields (see, for example, \cite{Feynman:1972:SMS}).

From operational point of view, the relation (\ref{eq:main}) implies that we cannot
control the friction kernels or friction coefficients without changing the bath-mediated interaction between the Brownian particles.  As a demonstration, if all the $\omega_{i,J}$ of the light particles are changed by a multiplicative factor $\lambda,$ i.e.   $\omega_{i,J}\mapsto \lambda\,  \omega_{i,J},$ then both  $K_{J,J'}(s)$ and  $U_{{b}}(X)$ should be changed to $\lambda^2 K_{J,J'}(\lambda s)$ and $\lambda^2 U_{{b}}(X),$ respectively. 

Especially about the work $W$ of operations, (\ref{eq:main}) implies that 
the work $W_{K}$ to change the off-diagonal friction kernel $K_{1,2}$
cannot be isolated from  the work  $W_{U_{}}$ to change the bath-mediated interaction potential,
$U_b.$ 
In the above solvable models, the total work, $ W= W_{K}+ W_{U_{}},$ to change the parameters, $\{\omega_{i,J}\},$ can be given as the Stieltjes integrals along the time-evolution of the whole degrees of freedom:
\beq  \label{eq:DW}
 W=\sum_{i}\sum_{J=1}^2\int_\Gamma
 \frac{\partial 
 H_{bB}}{\partial \omega_{i,J}}d\omega_{i,J}(t),
\eeq
where $\int_\Gamma$ indicates to integrate along the process where 
all the dynamical variables $p_{i'},\tilde{q}_{i'}$ and $X_J$ in the integrals evolves according to the system's dynamics under time dependent parameters $\{\omega_{i,J}\}$.
The operational inseparability of the work into $ W_{K}$ and $ W_{U_{}}$
 justifies the fact that, on the level of the stochastic energetics \cite{LNP}, we could not access the work to change the friction coefficients. 
On the microscopic level, however,  the above models allow to identify $ W_{K}$: 
First  $ W_{U_{}}$ is given by the above framework \cite{LNP}:
\eqn{\label{eq:DUBbB}
 W_{U_{}}=\sum_i  \sum_{J=1}^2  \int_\Gamma
 \frac{\partial U_{{b}}}{\partial \omega_{i,J}}\, d\omega_{i,J}(t),
}
because $U_0$ does not depend on $\omega_{i,J}.$
Combining (\ref{eq:DUBbB}) with (\ref{eq:DW}) as well as the identity (\ref{eq:Hb-bis}), 
the kinetic part of the work, $ W_{K},$ is found to be
\eqn{
 W_{K}= \sum_i  \sum_{J=1}^2 \int_\Gamma  \frac{\partial }{\partial \omega_{i,J}}
\inRbracket{\sum_{i'} 
\frac{m_{i'}\tilde{\omega}_{i'}^2}{2}\tilde{q}_{i'}^2
}\, d\omega_{i,J}(t),
}
where $\tilde{q}_i$ are defined in (\ref{eq:qtilde}). The result again shows that, unless we have an access to the microscopic fluctuations in the heat bath, $ W_{K}$ is not measurable.

In conclusion we propose, with supporting examples, that  a bath-mediated effective potential between the Brownian particles, $U_{{b}},$  should accompany the off-diagonal frictional memory kernel, $K_{1,2}(s),$ with a particular relation (\ref{eq:main}) due to the general sum rule of the linear response theory.
{This relation should be tested experimentally and/or numerically on the one hand, and the generalization to other models  \cite{nonMarkov-Bocquet-JSP1997,Hinch-exp3} should be explored on the other hand.}
For example, in  the reaction dynamics of protein molecules or of colloidal particles, non-local fluctuations of the solvent may play important roles both kinetically and statically.
The consciousness of the environment as a part of the whole system 
is important not only in the ecology but also at the micron- or nano-scale physics.

\begin{acknowledgments}
This work is supported by the Marie Curie Training Network NETADIS  (FP7, grant $290038$) 
for CDB. KS acknowledges Antoine Fruleux for fruitful discussions.
CDB and KS thank ICTP (Trieste, Italy) for providing with the opportunity to start the collaboration. 
\end{acknowledgments}

\bibliographystyle{apsrev4-1.bst}   
\bibliography{ken_LNP_sar}

           \end{document}